\documentclass{iopart}
\usepackage{amsmath,amssymb,graphicx}
\begin{document}
\title[The quantum Landau bound in systems with vanishing gap]{Achieving the Landau bound to precision of quantum thermometry 
in systems with vanishing gap}
\author{Matteo G. A. Paris}
\address{Quantum Technology Lab, Dipartimento di Fisica dell'Universit\`a 
degli Studi di Milano, I-20133 Milano, Italia}
\begin{abstract}
We address estimation of temperature for finite quantum systems at
thermal equilibrium and show that the Landau bound to precision $\delta
T^2 \propto T^2$, originally derived for a classical {\em not too small}
system being a portion of a large isolated system at thermal
equilibrium, may be also achieved by energy measurement in microscopic
{\em quantum} systems exhibiting vanishing gap as a function of some
control parameter.  On the contrary, for any quantum system with a
non-vanishing gap $\Delta$, precision of any temperature estimator
diverges as $\delta T^2 \gtrsim T^4 e^{\Delta/T}$.
\end{abstract}
\pacs{05.30.-d, 03.67.-a, 05.40.-a}
\date{\today}
\section{Introduction}
In the last decades, we have seen a constant improvement in the 
generation and control of engineered quantum systems, either to 
test quantum mechanics in a mesoscopic or macroscopic setting, 
or for the implementation of quantum-enhanced technologies.  
More recently, controlled quantum systems have become of interest 
to test and explore thermodynamics in the quantum regime, e.g. for
the characterization of work and energy statistics. Indeed
experiments in several optical and material systems 
have been suggested and implemented, with the
aim of understanding relaxation, thermalisation, and 
fluctuations properties in systems exhibiting explicit 
quantum features or being at the classical-quantum boundary
\cite{par77,ber85,par89,jar97,muk03,sei05,cuc92,klu93,ple98,Nie02,jac12,
gha14,ali14,pla14,bra15,bin15,bru15,bor15,esp15,ols15,pek15}.
\par
In this framework, it has become increasingly relevant to have a precise
determination of temperature for quantum systems
\cite{nmr0,nmr1,nmr7,nmr8,nmr9,nmr10,nmr11}, and to 
understand the ultimate bounds to precision in the estimation of 
temperature posed by quantum mechanics itself
\cite{bru06,sta10,bru11,bru12,mar13,hig13,cor14,meh15,jev15,jar15,adp15}.
The problem cannot be addressed in elementary terms
since, as a matter of fact, for a quantum system in equilibrium with a
thermal bath, there is no linear operator that acts as an observable for
temperature and we cannot write down any 
uncertainty relation involving temperature. In turn, this is
somehow connected with the fact that temperature, thought 
as a macroscopic manifestation of random energy exchanges between 
particles, {\em does not}, in fact, fluctuate for a system at thermal 
equilibrium. Therefore, in order to retain the operational definition 
of temperature, one is led to argue that although the temperature itself 
does not fluctuate, there will be fluctuations for any temperature 
estimate, which is based on the measurement of some proper observable 
of the systems, e.g. energy or population.
\par
This line of thought has been effectively pursued in classical 
statistical mechanics where, upon considering temperature as a function 
of the exact and fluctuating
values of the other state parameters, Landau and Lifshitz 
derived a
relation for the temperature fluctuations of a finite system
\cite{LanLi,Phi84}.  This is
given by $ \delta T^2 = T^2/C$ where $C$ is the heat capacity of the system
itself. In turn, this appears as a fundamental bound to the precision
of any temperature estimation. However, the relation has been
derived for a system which represents a small portion (but not too small) 
of a large, isolated, system in thermal equilibrium. Besides, it has been 
derived assuming the absence of any quantum fluctuation. Overall, 
its validity is thus questionable if the temperature is low enough 
or the system is known to exhibit quantum features \cite{fes87}.
\par
A first attempt to establish the Landau bound for finite quantum
systems has been pursued by inverting the energy dependence on 
temperature \cite{mah11}, however still assuming that the system is
large enough. Earlier, the concept of temperature fluctuations 
had caused longstanding controversies
\cite{Man6x,mcf73,Kit73,Kit88,Man89,Pro93,Chu92,Bol10}, 
which have not really solved to date, at least in fundamental 
terms \cite{uff99}.
In particular, it is not clear whether, and under which conditions, 
the Landau bound may be confirmed for quantum systems at low 
temperature (where quantum fluctuations become relevant), and 
whether the corresponding precision may be achieved in practice.
\section{The Landau bound in quantum systems with vanishing gap}
Our starting point is to recall that temperature is not an
observable and therefore its value should be estimated through an
indirect detection scheme, i.e. by measuring something else, say the 
observable $X$ on $M$ repeated preparations of the system, and then 
suitably processing the data sample  ${x_1,...x_M}$ in order to infer 
the value of temperature. 
The function $\hat T (x_1,...,x_M)$ is usually
referred to as an {\em estimator} and provide an operational
definition of temperature for the system under investigation.
However, as firstly noticed by Mandelbrot for closed systems 
\cite{Man6x,Man89}, different inference
strategies may be employed, e.g. by starting from different observables,
or just by using different estimators (say, the mean or the mode) 
on the same data sample, thus leading to different, and perfectly
acceptable, definitions of temperature,  In other words, temperature
for a thermodynamic system cannot uniquely defined, and no specific
definition can give rise to consensus. 
\par
On the other hand, we may give proper and unique definition to 
the notion of temperature fluctuations. In fact, the variance of 
{\em any} unbiased estimator of temperature is bounded by the 
Cramer-Rao theorem \cite{Cra46}, stating that 
\begin{equation}
\delta T^2 \geq \frac{1}{M F(T)}\,,
\label{crb}
\end{equation}
where $\delta T^2 \equiv \hbox{Var}(\hat T) = \langle(\hat T - T)^2\rangle$, 
$M$ is the number of repeated measurements and $F(T)$ is the 
so-called Fisher information, given by 
\begin{equation}
F(T)= \int\!\! dx\, p(x|T) \left[\partial_T \log p(x|T)\right]^2\,,
\label{fdef}
\end{equation}
being $X$ the quantity measured to infer the temperature and $p(x|T)$
the conditional distribution of its outcomes given the true, fixed,
value of temperature. The overall picture arising from the 
Cramer-Rao theorem is that the notion of temperature may be indeed 
imperfectly defined whereas, at the same time, the notion 
of temperature fluctuations may be given an unique meaning.
\par
We will fully exploit this approach to establish whether, and in which 
regimes, the Landau bound to thermometry may be established for 
quantum systems. To this aim, the crucial observation is that the 
quantum version of the Cramer-Rao theorem 
\cite{lqe1,lqe2,lqe3,lqe4,lqe5,lqe6} provides tools to 
individuate an {\em optimal} strategy to infer the value of temperature, 
i.e. to define a privileged observable related to temperature, which
allows one to determine temperature with the ultimate precision. This is done in 
two steps: i) find the observable that maximizes the Fisher 
information and ii) find an estimator that saturate the Cramer-Rao 
bound. The first step may be solved in a {\em system-independent} way, 
upon considering the observable defined by the spectral measure 
of the so-called {\em symmetric logarithmic derivative}, i.e. 
the self-adjont operator $L_T$ obtained by 
solving the Lyapunov-like equation 
\begin{equation}
\partial_T \varrho_T = \frac12 \left(L_T\varrho_T+\varrho_T L_T
\right)\,,\label{SLD-def}
\end{equation}
where $\varrho_T$ is the density operator of the system under
investigation. The second step is, in general, dependent on the system
under investigation, though general solutions may be found in the
asymptotic regime of large data samples, where Bayesian or
maximum-likelihood estimators are known to saturate the Cramer-Rao
bound. 
\par
In approaching the issue of temperature fluctuation, one often assumes
that the quantity that should be measured is the energy of the system.
The first thing we can prove using quantum estimation theory is 
that energy measurement is, in fact, the optimal one for any quantum 
systems.  The only assumption needed to prove this statement is that the 
system under examination may be described in the canonical ensemble.  
Let us denote by ${\cal H}$ the Hamiltonian of the quantum system under 
investigation and by ${\cal H} |e_n\rangle = E_n |e_n\rangle$ its eigenvalues 
and eigenvectors. At thermal equilibrium the density operator of the
system is 
\begin{equation}
\varrho_T = \frac1Z \exp\{-\beta {\cal H}\} = \frac1Z \sum_n e^{-\beta
E_n} |e_n\rangle\langle e_n|\,,
\label{rhoT}
\end{equation}
where $\beta=T^{-1}$, being the Boltzmann's constant set to one, and 
$Z=\hbox{Tr}[\exp\{-\beta {\cal H}\}]$ is the partition function of 
the system. Inserting Eq. (\ref{rhoT}) in Eq. (\ref{SLD-def}) we have
a solvable equation, leading to 
\begin{equation}\label{LT}
L_T = \sum_n\, \frac{E_n-\langle {\cal H}\rangle}{T^2}\,
|e_n\rangle\langle e_n|\,,
\end{equation}
where $\langle {\cal H}\rangle$ is the average energy of the system.
Eq. (\ref{LT}) shows that the optimal measurement is diagonal in the
Hamiltonian basis, i.e. it may be achieved by measuring the energy of
the system. The corresponding Fisher information is given by 
$F(T) \propto \langle \delta {\cal H}^2\rangle/T^4 = c_V(T)/T^2$, $c_V(T)$
being the specific heat at temperature $T$. 
In turn, this relation reveals that when the specific heat 
increases then the same happens to the Fisher information associated to 
temperature, e.g. temperature may be effectively estimated 
at the classical phase transitions with diverging specific heat
\cite{zan07,zan08}.
On the other hand, if the specific heat is bounded from above 
than precision of temperature estimation is bounded from below. 
\par 
We now proceed to investigate whether the Landau bound for quantum
systems holds also in the low temperature regime. In doing this, 
we analyze the microscopic origin of the behavior of the specific 
heat {\em without} making 
any assumptions of the size of the system. To this aim we assume 
that only the two lowest energy levels of the system are populated
(we are in the low temperature regime). The density operator may be
written as 
\begin{equation}\label{lowTrho}
\varrho_T = \frac1Z \left( |e_0\rangle\langle e_0| +
e^{-\Delta(\lambda)/T} |e_1\rangle\langle e_1| \right)\,,
\end{equation}
where the partition function reads as follows $Z=1+e^{-\Delta(\lambda)/T}$
and $\Delta\equiv\Delta(\lambda)=E_1-E_0$ is the energy gap between
the two levels. In writing (\ref{lowTrho}) we also assumed that 
the energy levels of the systems do depend on some external control
parameter $\lambda$, e.g. an internal coupling or an external field,
which may be exploited to tune the energy gap $\Delta(\lambda)$ between
the two levels.
Using Eq. (\ref{fdef}) and the fact that energy measurement is optimal, 
the Cramer-Rao bound (\ref{crb}) for temperature estimation says that 
the variance of any temperature estimator is bounded by 
$$
\delta T^2 \geq T^2 g(\Delta/T)
$$
where $$ g(x)= \frac{2}{x^2} (1 + \cosh x)\,.$$ 
\begin{figure}[h!]
\includegraphics[width=0.48\columnwidth]{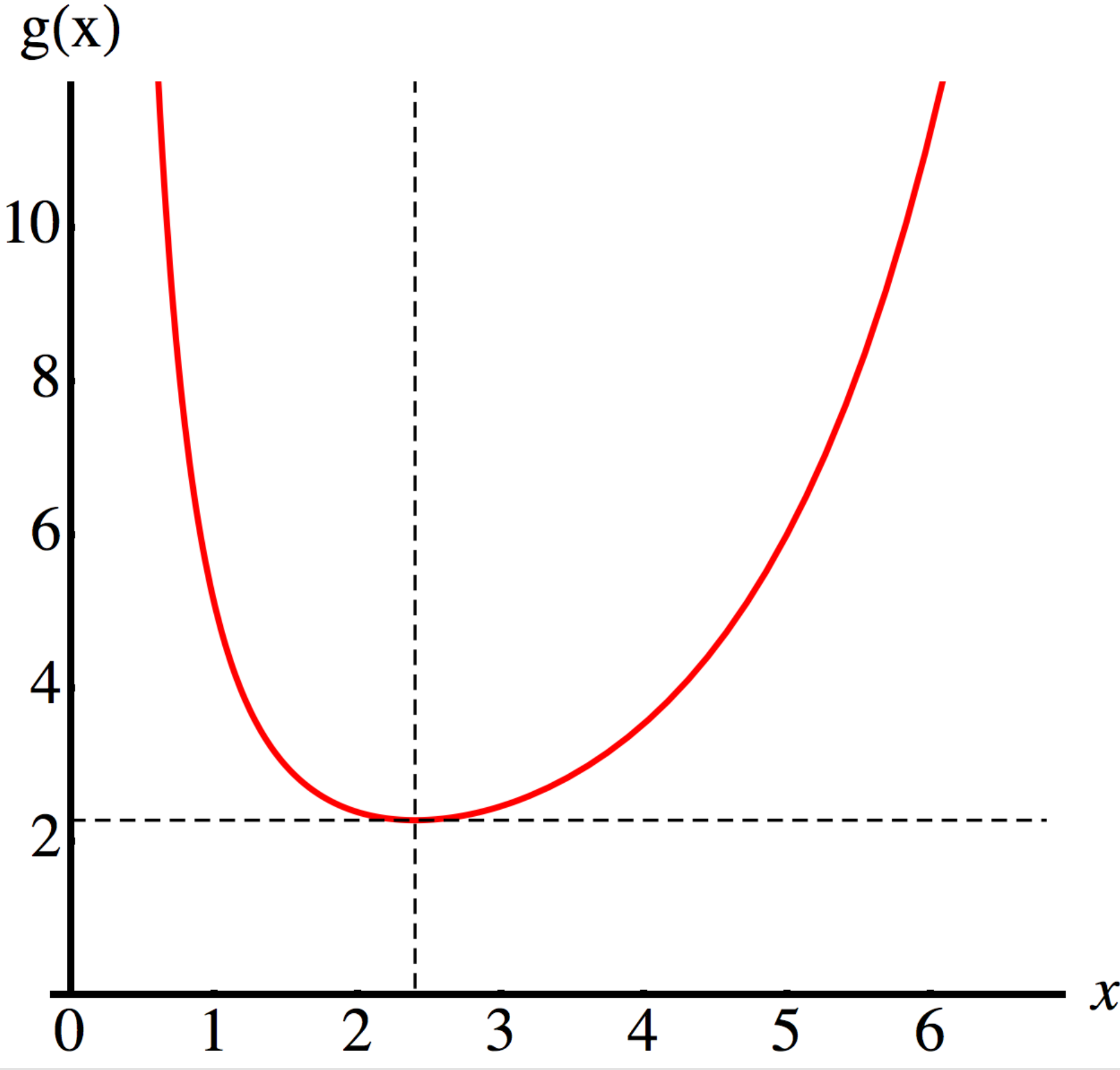}
\includegraphics[width=0.48\columnwidth]{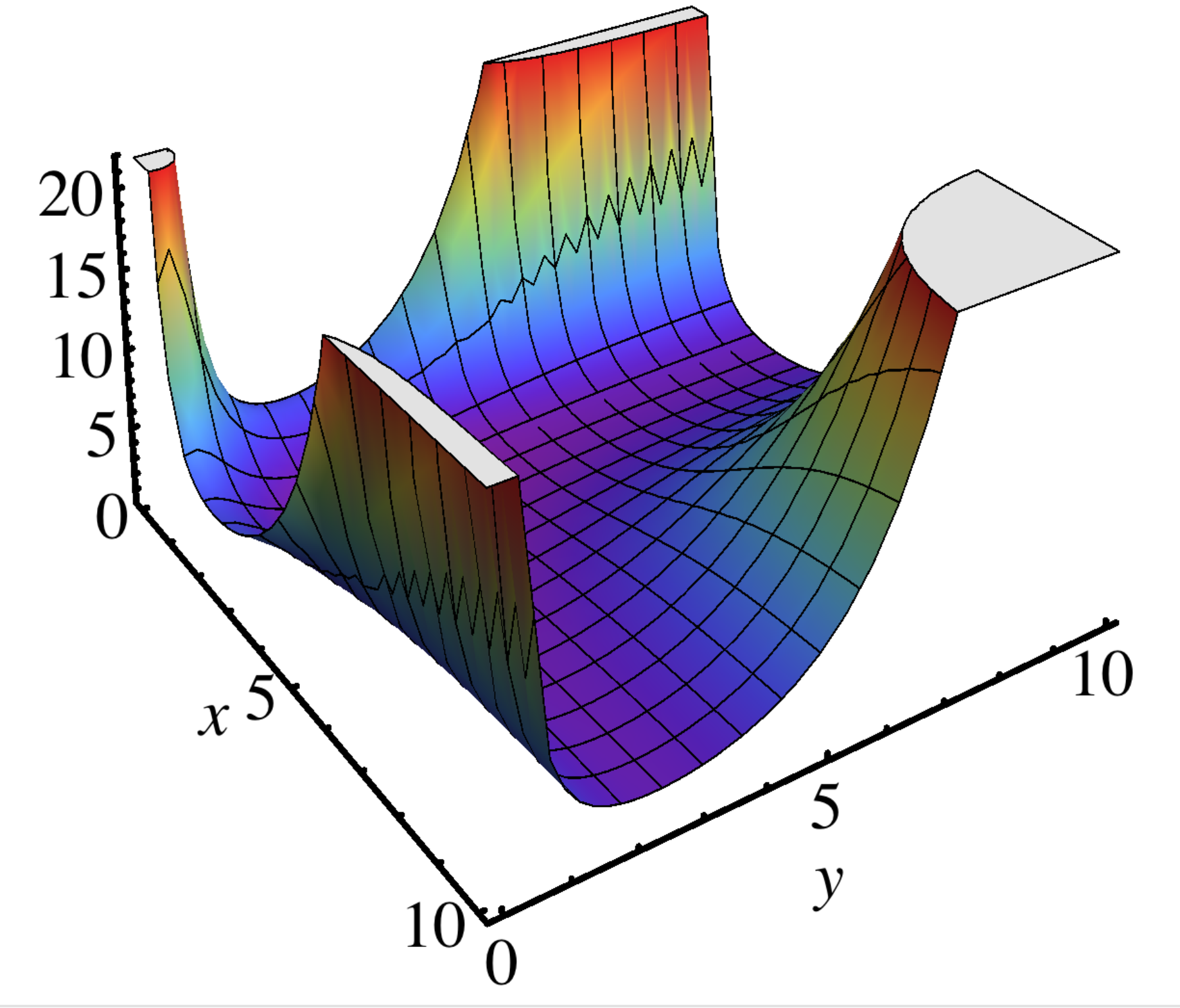}
\caption{\label{f:g} The functions $g(x)$ and $h(x,y)$ 
governing the variance of temperature estimators in the 
low temperature regime.}
\end{figure}\par
The function $g(x)$ is depicted in the left panel of Fig. \ref{f:g}. 
It diverges as 
$e^x/x^2$ for $x\rightarrow \infty$ and as $4/x^2$ for $x\rightarrow 0$, 
whereas it shows a minimum $g(x_m)
\simeq 2.27$ located at $x_m\simeq
2.4$. It follows that in systems where the gap $\Delta$ may be tuned
to arbitrarily small values by tuning the external control $\lambda$, 
such that $\Delta/T\simeq x_m$ remains finite, optimal estimation 
of temperature with precision at the Landau bound $\Delta T^2 
\propto T^2$ may be achieved by measuring energy and a suitable 
data processing. On the contrary, in system where the gap has a 
minimum, temperature may be estimated efficiently only down to 
a threshold, below which the variance of any estimators starts 
to increase as 
$$
\delta T^2 \gtrsim T^4 e^{\Delta/T}\,.
$$
The above results are valid for arbitrarily small systems at 
low temperature and do not depend on the specific structure 
of the system Hamiltonian, nor on the size of the system. 
The only requirement is that the
system exhibits vanishing gap between its lowest energy levels
as a function of some external control parameter. Results are 
also independent on any specific features of the two-level 
approximation, assuming that a gap 
above the first excited level is present in order to make sense 
of the two-level description.
The range of temperature where the results holds corresponds
to the range of validity of the two-level approximation, roughly 
speaking $T$ of the order of the gap above 
the first excited level. Results are however robust against this
parameter. To confirm 
this statement, let us consider a three-level approximation where 
$$\varrho_T = \frac1Z \left( |e_0\rangle\langle e_0| +
e^{-\Delta_1(\lambda)/T} |e_1\rangle\langle e_1| +  
e^{-\Delta_2(\lambda)/T} |e_2\rangle\langle e_2| \right)\,,$$ and
$Z=1+e^{-\Delta_1(\lambda)/T}+e^{-\Delta_2(\lambda)/T}$,
$\Delta_k=E_k-E_0$. The resulting Cramer-Rao bound, for energy
measurement, is given by $$\Delta T^2 \geq T^2\,
h(\frac{\Delta_1}{T},\frac{\Delta_2}{T})\,,$$
where $$ h(x,y) = \frac{e^{-x-y}(e^x+e^y+e^{x+y})^2}{(1+e^y)x^2
-2xy+(1+e^x)y^2}\,.$$ 
The function $h(x,y)$ is depicted in right left panel of Fig. \ref{f:g}.
It is symmetric and shows a global minimum $h_m\simeq 1.31$ located at
$x_h=y_h\simeq2.66$. It also shows local minima at $x_m\simeq2.4$ for
increasing $y$. Upon  tuning the gap $\Delta_1$ to arbitrarily small
values such that $\Delta/T\simeq x_m$ remains finite, we have
$y=\Delta_2/T \rightarrow \infty$. On the other hand, $h(x,y)$
approaches $g(x)$ for
increasing $y$ and we are thus smoothly back to the two-level case.
\subsection{Remarks}
Estimation theory has been also used to address properties
of thermometers, rather than intrinsic properties of the system
under investigation. In particular, the role of the 
numbers $N$ of particles has been analyzed, showing that 
performing energy measurement on a non thermalizing 
thermometer made of two-level atoms
allows one to improve scaling of precision from $N^{-1/2}$ to $N^{-1}$
\cite{sta10}.  The analysis has been also extended to thermometer made
of multilevel atoms \cite{cor14}, either fully or partly thermalizing,
showing that the sensitivity grows significantly with the number of
levels, with the optimization over their energy spectrum playing a
crucial role.  We emphasize that this results pertain to properties of
quantum thermometers. i.e. quantum systems used to probe the temperature
of an external bath, whereas our focus has been on establishing
intrinsic bounds to precision, thus providing benchmarks to assess any
detection scheme.
It should be also mentioned that upon employing arguments
similar to those used in \cite{sta10} the analysis of the previous
Section may be extended to degenerate systems, where $\Delta$ now
represents the average energy per particle.
\par
Another remark concerns a possible, alternative, explanation introduced
to account for fluctuations in temperature measurements.  The argument
is based on the idea that an intrinsic distribution of temperatures may
exists, which is consistent with a given thermodynamic state, without
implying dynamical fluctuations of the temperature(s) themselves. The
argument is usually referred to as the polythermal ensemble hypothesis
\cite{Phi84}, and it has been somehow criticized \cite{Pro93} since it
requires more hypothesis than just assuming the canonical ensemble.
\par
Finally, it should be mentioned the use of a Hamiltonian 
control parameter to improve thermometric strategies has been 
already implemented experimentally, e. g. for strongly interacting 
Fermi gases \cite{Luo07,Luo09}.
\section{Conclusions}
In the recent years, schemes for temperature estimation involving the
interaction of the system with an individual quantum probe have received
attention
\cite{sta10,bru11,bru12,mar13,hig13,cor14,meh15,jev15,rai15,emm13}
mostly because they provide temperature estimate by adding the minimal
disturbance. Our results, which have been obtained with basically no
assumptions on the structure of the system under investigation and on
the measurement performed to extract information, provide a general
benchmark to assess this schemes, and to design effective thermometers
for quantum systems.
\par
Our results also provide a framework to reconcile the different
approaches to temperature fluctuations. As a matter of fact, temperature
itself does not fluctuate, however, there are fluctuations for the
temperature estimate based on any indirect measurement. In other words,
temperature is a classical parameter which do not correspond to a 
quantum observable and estimation of
temperature necessarily involves the measurement of another quantity,
corresponding to a proper observable. In turn, quantumness in
temperature estimation is in the measurement stage and in
the nature of fluctuations of the measured observable. 
\par
The optimal strategy to estimate temperature of a small quantum system
turns out to be measuring the energy of the system and suitably process
data, e.g. by Bayesian analysis \cite{b1,b2}, in order to achieve the
Cramer-Rao bound to precision. In this way, we have shown that the
classical Landau bound to precision is recovered, in the low temperature
regime, for systems exhibiting a vanishing gap as a function of some
control parameter.  On the contrary, in systems with a non-vanishing gap
$\Delta$ between the lowest energy levels, temperature may be
effectively estimated only down to a threshold, below which the variance
of any estimator starts to increase as $\delta T^2 \gtrsim T^4
e^{\Delta/T}$.  Notice that this is true independently on the use of an
external ancillary  system to probe the temperature of the system under
investigation. In other words, rather that being a property of the
"thermometer"  (i.e. of the chosen ancillary system and of the probing
interaction scheme),  the ultimate precision in temperature estimation
is an intrinsic property of the quantum system itself.  Our analysis
shows the optimality of quantum thermometry based on energy
measurements, and provides quantum benchmarks for high precision
temperature measurement, as well as an efficient operational
quantification of temperature for quantum mechanical systems lying
arbitrary close to their ground state.  
\ack
This work has been supported by EU through the Collaborative Projects
QuProCS (Grant Agreement 641277) and by UniMI through the H2020
Transition Grant 14-6-3008000-625. The author thanks Antonio Mandarino
and Giulio Salvatori for discussions in the initial stage of this work
and Valentina De Renzi for discussions about gapped systems. The author 
also thanks the anonymous Referees for their comments. This work
is dedicated to the memory of R. F. Antoni.
\section*{References}

\end{document}